\documentclass[aps,twocolumn, superscriptaddress]{revtex4-1}

\usepackage{float}
\usepackage{amssymb}
\usepackage{amsmath}
\usepackage{upgreek}
\usepackage{ulem}
\usepackage[shortlabels]{enumitem}

\usepackage{graphicx}

\usepackage{color}

\begin{document}

\title{Quantitative analysis of the prediction performance of a Convolutional Neural Network evaluating the surface elastic energy of a strained film}

\author{Luis Martín-Encinar}
\altaffiliation{Contributed equally}
\affiliation{Dpto. de Electricidad y Electrónica, E.T.S.I. de Telecomunicación, Universidad de Valladolid, 47011, Valladolid, Spain}

\author{Daniele Lanzoni}
\altaffiliation{Contributed equally}
\affiliation{L-NESS and Dept. of Materials Science, Universit\`{a} di Milano-Bicocca, 20125, Milano, Italy}

\author{Andrea Fantasia}
\affiliation{L-NESS and Dept. of Materials Science, Universit\`{a} di Milano-Bicocca, 20125, Milano, Italy}

\author{Fabrizio Rovaris}
\affiliation{L-NESS and Dept. of Materials Science, Universit\`{a} di Milano-Bicocca, 20125, Milano, Italy}

\author{Roberto Bergamaschini}
\email{roberto.bergamaschini@unimib.it}
\affiliation{L-NESS and Dept. of Materials Science, Universit\`{a} di Milano-Bicocca, 20125, Milano, Italy}

\author{Francesco Montalenti}
\affiliation{L-NESS and Dept. of Materials Science, Universit\`{a} di Milano-Bicocca, 20125, Milano, Italy}

\date{\today}

\begin{abstract}  
A Deep Learning approach is devised to estimate the elastic energy density $\rho$ at the free surface of an undulated stressed film. About 190000 arbitrary surface profiles $h(x)$ are randomly generated by Perlin noise and paired with the corresponding elastic energy density profiles $\rho(x)$, computed by a semi-analytical Green's function approximation, suitable for small-slope morphologies. The resulting dataset and smaller subsets of it are used for the training of a Fully Convolutional Neural Network. The trained models are shown to return quantitative predictions of $\rho$, not only in terms of convergence of the loss function during training, but also in validation and testing, with better results in the case of the larger dataset. Extensive tests are performed to assess the generalization capability of the Neural Network model when applied to profiles with localized features or assigned geometries not included in the original dataset. Moreover, its possible exploitation on domain sizes beyond the one used in the training is also analyzed in-depth. The conditions providing a one-to-one reproduction of the "ground-truth $\rho(x)$ profiles computed by the Green's approximation are highlighted along with critical cases. The accuracy and robustness of the deep-learned $\rho(x)$ are further demonstrated in the time-integration of surface evolution problems described by simple partial differential equations of evaporation/condensation and surface diffusion.
\end{abstract}

\maketitle

\section{Introduction}
In the last decade, applications of Machine Learning approaches to Materials Science have grown exponentially, blown by the development of highly parallelized computational architectures, especially based on GPUs. Neural Networks (NN)~\cite{goodfellowdeep2016, Mehta2019PhysRep}, already used for fitting interatomic potentials \cite{Behler2007PRL, Bartok2010PRL}, are now finding applications on a wide variety of problems, ranging from experimental data analysis to numerical simulations~\cite{venderley2018machine, salmenjoki_machine_2018, Jha2018grain_orientation, Yang2021dopants, ge2024silicon}, thanks to their flexibility and capabilities of approximating functions with arbitrary precision~\cite{cybenko1989approximation}. The use of NNs for accelerating computationally demanding tasks allows to achieve the desired outputs at a fraction of the time requested by conventional methods and enables the study over spatial- and/or temporal-scales that would not be otherwise accessible.

Recent studies have shown how this concept can be proficiently exploited not just on the atomistic scale but also in the continuum~\cite{raissi2019physics, Fulton2019CGF, MontesdeOcaZapiain2021npj, Yang2021Patterns, Lanzoni2022PRM, Zeqing2023APLML}, proposing NN workflows to completely or partially surrogate the numerical solution of Partial Differential Equations governing physical problems. In this context, Convolutional NNs (CNN) are particularly convenient for the approximation of differential operators, as they intrinsically encode spatial correlations and symmetries.

Here we apply a CNN to replace the solution of the mechanical equilibrium problem for a strained film. More precisely, the prototypical case of a Ge film subject to a 4\% compressive strain, mimicking the epitaxial misfit with a hypothetical Si(001) substrate, will be considered, with no loss of generality. For the sake of simplicity, we here target to predict by NN just the map of the elastic energy density $\rho$ at the film surface but the proposed approach could be straightforwardly applied to any other scalar property as well as to the components of the displacement vector or of the stress/strain tensors. Given that the present work aims at providing a quantitative evaluation of the efficacy of the proposed approach, hundreds-of-thousands of examples are needed for training and testing the NN, each one requiring the solution of the elastic problem to extract the ``true" $\rho$ profile. To this goal, we find it convenient to exploit a semi-analytical Green's function approximation (GA) \cite{landau,tersoff1994,Zinovyev} to evaluate the strain state of a non-planar, low-aspect-ratio, film in two dimensions (2D). The understanding on the NN behavior acquired after this study, allowing for the identification of the optimal feature set to be considered for proficient NN training, poses the basis for application-oriented developments, using more-accurate, yet time-consuming, calculation methods. An extension using a Finite-Element Method solver in place of GA is the object of an upcoming publication.

The use of a Fully-Convolutional NN architecture~\cite{long2015fully} is particularly appealing in the case at hand, as it allows for a convenient encoding of physical symmetries by construction~\cite{Lanzoni2022PRM}. As in all ML approaches, the choice of good inductive biases is also critical to reducing the number of examples required to learn an accurate model~\cite{goodfellowdeep2016, Mehta2019PhysRep}. Still, the question of how big a dataset should be to provide an effective approximation is an important one for this class of methods, particularly in cases in which data acquisition is expensive or time-consuming. For this reason, we train models with different dataset sizes and compare performances on suitable validation and test sets.

Once the NN model has been trained, the next step involves an extensive inspection of its capabilities, including assessing its ability to generalize unseen examples, comparing its predictions against ground-truth solutions based on GA, and evaluating its reliability in both extrapolation and interpolation scenarios. By this comprehensive examination, we may discern the strengths and weaknesses of the model in order to confidently apply it. Moreover, the NN predictions can be exploited in more complex tasks, such as the time-integration of surface evolution equations\cite{FreundBOOK2003,Mullins1957JAP,BergamaschiniAPX2016}, where $\rho$ represents the elastic driving force for the material redistribution. However, while accurate predictions are achieved for static configurations, relying on the deep-learned $\rho$ at each of the possibly millions of timesteps of an evolution simulation implies even stronger constraints to warrant stable and consistent behavior. Therefore, testing time-evolution is critical for validating the quality of NN predictions.

The paper is organized as follows. First, in the Sect.~\ref{sec::methods} we describe the study methodology, including the evaluation of $\rho$ in the Green's function approach, the details of the CNN structure, and the dataset generation. Then, in the Sect.~\ref{sec::results} we discuss the training and validation procedure, and investigate the performance of the NN to provide quantitative predictions of $\rho$ against the GA ground-truth solution. The strengths and limits of the NN, as well as its generalization capabilities, are analyzed by an extensive testing. A few applications to the solution of well-known time-dependent problems are finally reported to demonstrate the robustness of the proposed approach, yielding one-to-one match of the GA-based ground-thruth simulations, and its extensibility to large-scale and long-time. Last, conclusions and perspectives are discussed.

\section{Methods}\label{sec::methods}

\subsection{Elastic energy density and Green's function approximation}\label{sec::phys_model}
In the present work, we investigate the strain state of a semi-infinite film, described by a 2D profile function $y=h(x)$. For the sake of simplicity, the film is described as an elastically isotropic medium, although extensions to non-isotropic linear elasticity could be considered using the same framework. As we refer to a Ge film, we take $Y$=103~GPa for the Young modulus and $\nu$=0.26 for the Poisson ratio.

The strain state of the film is found by solving the mechanical equilibrium equations, in the linear elasticity regime~\cite{landau}:
\begin{align}\label{eq::elastic_problem}
    & \nabla \cdot {\bf \sigma} = \nabla \cdot \left[{\bf C} {\bf\colon} ({\bf\varepsilon} - {\bf\varepsilon^*}) \right]= {\bf 0}\,,\\
    & {\bf \sigma} \cdot {\hat{\bf n}}=0    \qquad \text{at free-surfaces}
\end{align}
where ${\bf \sigma}$ and ${\bf \varepsilon}$ are the stress and strain tensors, ${\bf \varepsilon^*}=4\%{\bf I}$ is the (diagonal) eigenstrain tensor, accounting for the nonelastic deformation state of the unrelaxed film \cite{eshelby1957determination, mura2013micromechanics}, and ${\bf C}$ is the tensor of elastic constants. For general free-surface profiles, eq.~\eqref{eq::elastic_problem} has no closed-form solution. However, for small-slopes a Green's function approximation \cite{landau,tersoff1994, Zinovyev} can be exploited. In particular, it is found that the in-plane $xx$ strain component can be expressed as:
\begin{equation}\label{eq::epsxx}
\varepsilon_{xx}(x) = - \frac{Y\varepsilon^* }{1-\nu^2} \int G_{xx}(x-x')h(x) dx'\,.
\end{equation}
$G_{xx}$ is the Green’s function for the strain field in a half-space, approximated by a Lorentzian function with cutoff $b$ to avoid the divergence at $x\rightarrow 0$:
\begin{equation}\label{eq::gxx}
G_{xx}(x) = \frac{2(1-\nu^2)}{\pi Y}\frac{1}{x^2}\approx \frac{2(1-\nu^2)}{\pi Y}\frac{1}{x^2+b^2}\,.
\end{equation}
The convolution integral in eq.~\eqref{eq::epsxx} can then be solved by means of Fourier transform as
\begin{equation}
    {\hat{\varepsilon}}_{xx}(q) = 2 \varepsilon^* \frac{e^{-b|q|}-1}{b} \hat{h}(q)\,.
\end{equation}
Once $\varepsilon_{xx}$ is recovered by Fourier anti-transformation, one can also straightforwardly evaluate the out-of-plane $yy$ strain component as $\varepsilon_{yy}=(\varepsilon^*-\nu \varepsilon_{xx})/(1-\nu)$.

The elastic energy density $\rho$ at each point $x$ along the film surface, which is the quantity of interest for the present study, is then:
\begin{equation}\label{eq::rho_gen}
  \rho(x)=\frac{1}{2} {\bf \sigma}{\bf \colon}({\bf \varepsilon}-{\bf \varepsilon^*})\approx \frac{Y}{2(1-\nu^2)}(\varepsilon_{xx}(x)-\varepsilon^*)^2\,.
\end{equation}

\subsection{Neural Network}\label{sec::NN}
Both $h$ and $\rho$ are represented by arrays containing values on a fixed, uniform mesh of $N_x$ points. To learn the approximate mapping between the free-surface profile and the corresponding elastic energy density, a specialized Fully Convolutional Neural Network~\cite{long2015fully} architecture has been implemented within the PyTorch framework~\cite{paszkepytorch2019}. As a broad definition, this class of networks is composed by stacking only convolutions and point-wise non-linearities, thus making it able to process inputs of arbitrary size. This property, particularly convenient in tasks such as image segmentation or image-image translation~\cite{long2015fully, choy2019fully, lata2019image}, can be leveraged in the present setting to apply the learned approximation on computational cells with a different size than the one presented to the NN at training time. This point, along with applicability limitations, will be analyzed in  Sects.~\ref{sec::sin_gauss}, \ref{sec::large_domain_statics} and ~\ref{sec::dynamics}, where the model is applied to domains of different sizes than those used in training. Besides this multiple-scales generalization possibility, another main advantage is that Fully-Convolutional NN can be easily modified to comply with physical symmetries by construction, effectively reducing the number of required examples and helping to control generalization error.

Translation equivariance is satisfied by construction by the fully-convolutional structure, as can be directly confirmed by the fact that the NN commutes with profile translation operators~\cite{goodfellowdeep2016}. In more practical terms, this means that if the profile $h(x)$ undergoes a horizontal shift $a$ transformation $h(x) \rightarrow h(x+a)$, then the predicted $\rho(x)$, will be correspondingly shifted by construction. Consistently with the Green's function approximation exploiting Fourier transforms, the NN adopted circular padding~\cite{Schubert2019IEEE, Kayhan2020CVF}, effectively implementing built-in periodic boundary conditions. Additionally, in the present case of a semi-infinite material, the elastic energy density at the surface is invariant with respect to vertical shifts. The NN prediction $\rho$ should therefore satisfy the condition $\rho(h(x)) = \rho( h(x) + b )$, for any vertical shift $b$ in the surface profile. A simple way to implement this symmetry, which has been exploited in the present work, is the subtraction of the average of the input profile every time $\rho$ is calculated. A last equivariance property of the profile-elastic energy mapping, valid for the present assumption of isotropic elastic constants, is mirror symmetry. If the profile is reflected with respect to $x$ coordinates, i.e. $h(x) \rightarrow h(-x)$, then $\rho(x)$ must be reflected in the same way. This symmetry was directly implemented in the NN structure by imposing that convolution kernels are invariant with respect to reflections: given a convolution kernel of size $N$ $[k_1,k_2,...,k_N]$, then the kernel $\frac{1}{2}[k_1+k_N, k_2+k_{N-1},...,k_N+k_1]$ is invariant to reflections by construction.

The standard mean-squared error for regression tasks is selected as a loss function. Given the predicted elastic energy density $\rho^\text{NN}$ and the ground-truth values computed by GA, $\rho^\text{GA}$ it reads:
\begin{equation} \label{eq::loss}
    \mathcal{L}(\vartheta) = \frac{1}{N_\text{TS} N_x} \sum_{i=1}^{N_\text{TS}} \sum_{j=1}^{N_x} (\rho^\text{NN}_j(h_i| \vartheta) - \rho^\text{GA}_j(h_i))^2\,,
\end{equation}
where the index $i$ runs on training examples (being $N_\text{TS}$ their total number), $j$ enumerates collocation points and $\vartheta$ are the NN parameters. Notice that the $\rho_j^\text{NN}$ value at the $j$-th collocation point is a function of the whole profile $h_i$.

We also report for the technical reader the specific architecture that in our tests produced the best results. The Network is built by stacking multiple simple-convolution/non-linearity (hyperbolic tangent activation was used) blocks, as it is standard in Deep Learning architectures. No skip connections are present. Convolutions encompass 20 different kernels with size 21. A total of 5 such basic blocks are used. With this structure, the total number of parameters $\theta$ to be learned is 26121, although symmetry constraints approximately halve the number of independent ones. Additional details and the specific code implementation can be found at \url{https://github.com/mosegroup/nn4surf}.

\subsection{Dataset generation}\label{sec::dataset}
The generation of high-quality datasets plays a key role in training, thus the creation of surface profiles that represent a wide spectrum of qualitatively distinct features is essential. To automatically generate arbitrary profiles of sufficient variability, we considered a kind of gradient noise known as Perlin noise \cite{Perlin1985287}. It inherently offers the advantage of producing random profiles with specified correlation lengths, ensuring that the generated dataset is not only diverse but also representative of possible surface morphologies. The Python project from Ref.~\cite{PythonPerlinNoise} has been used to conveniently generate profiles for training set instances.

Each of the generated $h(x)$ profiles is then associated with the corresponding elastic energy density $\rho(x)$ calculated by the Green's function approximation discussed in Sect.~\ref{sec::phys_model}.

Based on the well-known Asaro-Tiller-Grinfeld (ATG) theory \cite{Asaro1972MetalTrans,Grinfel1986SPD,Grinfeld1993JNS,SrolovitzAM1989}, the characteristic length scale for the morphology of strained-Ge films is expected to be of the order of a few tens of nms. A domain size of $L$=100~nm is then assumed sufficient to represent the main traits of the surface profiles. Then, in order to comply with the small-slope prescription of the GA approach, the peak-to-valley height of the generated profiles was restricted in the range [0.001,8]~nm, taking random values, equally spaced on a logarithmic scale.

A one-dimensional grid with a fixed $\delta x$=1~nm spacing is used for the profile discretization and each data point is identified as a pixel in the NN structure. A dataset of approximately 190000 examples, available from the repository~\cite{repository}, was been generated. Each instance consists of the list of 100 (scalar) $(h_j, \rho_j)$-pairs for the discretized profiles.

\section{Results and discussion}\label{sec::results}
In this section, we analyze the capability of the NN to provide one-to-one predictions of the elastic energy density $\rho(x)$ associated to arbitrary surface profiles $h(x)$. To quantify the accuracy of the NN predictions against the GA ground-truth, we consider both the local prediction error
\begin{equation}
  \delta_\rho(x)=\rho^\text{NN}(x)-\rho^\text{GA}(x)\,,
\end{equation}
evaluated at each point $x=x_i$ of the domain discretization, and the normalized root-mean-squared error:
\begin{equation}\label{eq::nrmseeq}
   \sigma_\rho = \frac{\sqrt{\dfrac{1}{N_x}\sum_i^{N_x} \left[\rho^\text{NN}(x_i) - \rho^\text{GA}(x_i)\right]^2}}{\text{max}(\rho^\text{GA})-\text{min}(\rho^\text{GA})} \,,
\end{equation}
where $i$ indexes each of the $N_x$ collocation points in the profile discretization. A 2\% value of $\sigma_\rho$ is identified as the limiting value to consider the NN predictions as a one-to-one reproduction of the ground-truth.

\subsection{Training, validation and testing}\label{sec::results_trainvalidtest}
Training is performed using the standard PyTorch implementation of the Adam optimizer~\cite{ paszkepytorch2019,kingmaadam2014}, with a learning rate of 1$\times$10$^{-5}$. Three different dataset sizes $N_\text{set}$ have been considered: the full set of 190000 examples and two subsets made of 70000 and 15000 examples respectively, which allows to understand the effect of the number of examples on the quality of the learned approximation. The dataset elements are randomly distributed into training and validation sets with a 70-30\% proportion. The minimization of the loss function defined in eq.~\eqref{eq::loss} is performed using a batch size of 512 and was halted once the average slope of the validation loss curve falls below 10$^{-4}$ (i.e. when it would take 10$^4$ iterations to reduce the loss by one order of magnitude).

\begin{figure}[b!]
 \centering
   \includegraphics[width=\columnwidth]{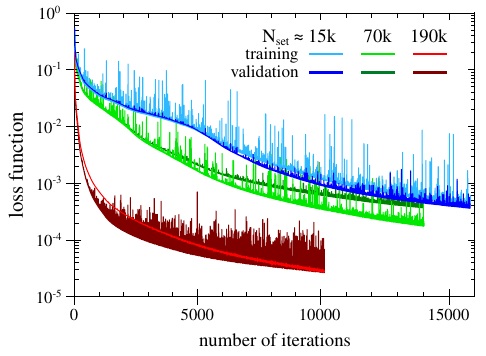}
    \caption{Logscale-plot of the training and validation loss values as a function of the number of training iterations for three NN models with dataset size $N_\text{set}\approx$15000, 70000 and 190000.}
    \label{fig::trainvalid}
\end{figure}

As customary~\cite{bishop2006pattern, goodfellowdeep2016, Mehta2019PhysRep}, the first quality assessment for a trained model is the comparison of the loss function as calculated on the training and validation set. The results for the three trained models are reported in Fig.~\ref{fig::trainvalid}. In all models, the losses steadily decrease during the training epochs without signs of overfitting. As expected, the model trained on the largest dataset outperforms the smaller ones, at least in terms of training/validation losses, reaching an $\mathcal{L}$ value at least one order of magnitude lower.

\begin{figure*}[htb!]
 \centering
   \includegraphics[width=\textwidth]{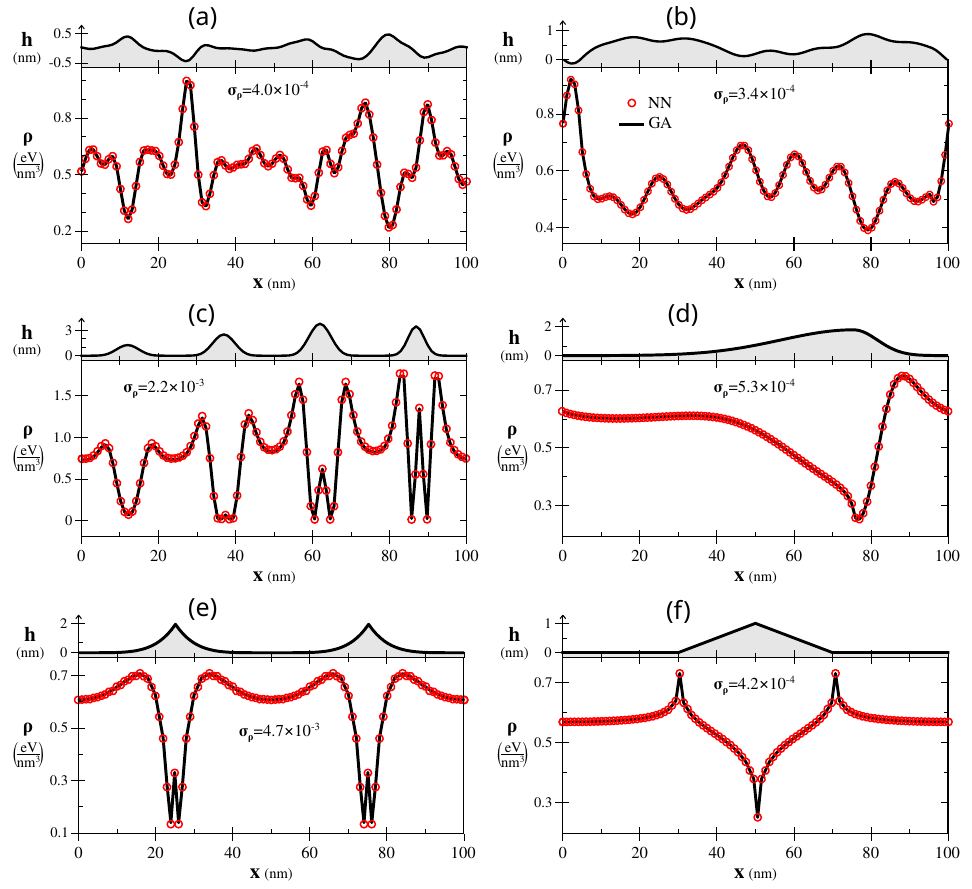}
    \caption{Testing of the trained NN model on different profiles $h(x)$: (a) a Perlin noise profile not included in the training set; (b) a profile made of random Fourier components; (c) a series of gaussian peaks; (d) an asymmetric peak; (e) a series of cusps; (f) a triangular peak. The corresponding elastic energy density $\rho(x)$ profiles predicted by the NN (red circles) and calculated by GA (solid black line) are superimposed. The prediction error $\sigma_\rho$ is reported for each case.}
    \label{fig::testingcases}
\end{figure*}

To certify the absence of biases in the NN model and to check extrapolation limits, several tests have been conducted on additional profiles not included in the training/validation set (an extensive selection is available in the repository~\cite{repository}). In particular, we identify six different classes of profiles, exemplified in Fig.~\ref{fig::testingcases}, useful to test the NN predictivity:
\begin{enumerate}[(a)]
    \item Perlin noise profiles, conformal to the ones used for the NN training but never exposed to it;
    \item profiles generated by random Fourier synthesis of cosines with $\lambda$ in the range [15,100]~nm, providing smooth profiles generated on a different basis than the one of training;
    \item combinations of non-overlapping gaussian peaks of different heights and widths, so to test the NN against localized features;
    \item asymmetric, single peaks of different skewness and heights, made by joining two gaussian halves of different widths, so to test the effect of symmetry in the NN predictions;
    \item series of cusps of different slopes with different heights and widths, to test the effect of discontinuous slope;
    \item profiles made of a single triangular peak of different height-to-base aspect ratios at the center of a flat domain, so to test polygonal geometries with straight angles.
\end{enumerate}
The analysis of profiles different from Perlin noise is meant to inspect the generalization capabilities of the trained NN-models on qualitatively different geometries, never analyzed in the training procedure. This is critical, as the non-linear nature of the NN could lead to uncontrolled approximation errors in extrapolation regimes, even if good performances were achieved in training and validation examples.

In Fig.~\ref{fig::testingcases} a comparison between the $\rho(x)$ profile calculated by GA and its prediction by the $N_\text{set}\approx$190000 NN model is reported for a case of each class of $h(x)$ profiles. A quantitative agreement is found for all cases as indicated by the $\sigma_\rho$ values, of the order of 10$^{-4}$-10$^{-3}$. Notably, the NN is found to correctly reproduce even the sharp peaks at the singular points in case (e) and (f).

\begin{figure*}[htb!]
 \centering
   \includegraphics[width=\textwidth]{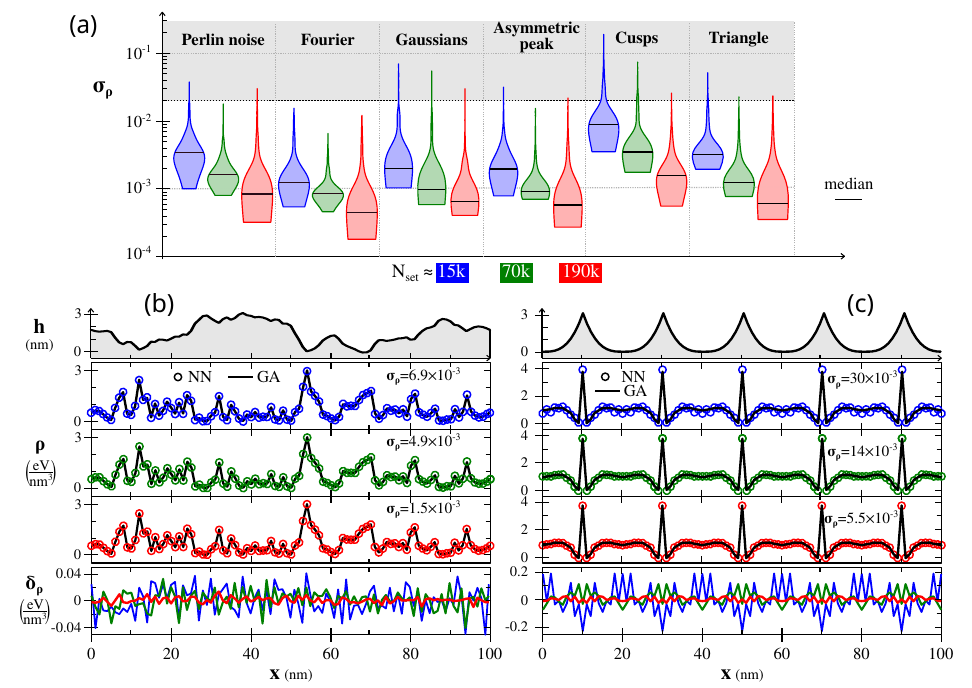}
    \caption{(a) Distribution of prediction errors $\sigma_\rho$ by violin plots for the three NN models with dataset size $N_\text{set}\approx$15000, 70000 and 190000, on the cases of a test set with profiles generated by variants of Perlin noise, superpositions of random Fourier components, series of gaussian peaks, an asymmetric peak, series of cusps and a triangular profile. (b, c) Comparison of the $\rho(x)$ profiles predicted by the three NN models (circles) with respect to the true GA solution (solid line) for (b) a Perlin-noise profile not included in the dataset and (c) a series of cusps. The corresponding local prediction errors $\delta_\rho$ are reported along with $\sigma_\rho$ values. }
    \label{fig::size}
\end{figure*}

In order to perform a more systematic analysis and assess the difference in predictivity of the three NN models trained on different dataset sizes, a testing set was then built by taking 1000 random profiles for each class, setting peak-to-valley heights equidistributed in log-scale in the same range as in the training set, to comply with the small-slope constraint and the training example range.

The distribution of $\sigma_\rho$ prediction errors returned by the three trained NN models on all cases in the six classes of the testing set is reported by violin plots in Fig.~\ref{fig::size}(a). Notably, all three trained models show satisfactory performances with median values of the order of 10$^{-3}$ or less and just a few cases exceeding the 0.02 acceptance threshold. This provides an a-posteriori justification for the choice of using Perlin noise as a generator of sufficiently various surface profiles, giving enough information to the NN model for generalizing to qualitatively different geometries. As observed for the training and validation losses, it is clear that the NN performances improve by taking a larger dataset but it is remarkable how the smallest NN-model, trained on just $\approx$8\% of the full dataset (i.e. $N_\text{set}\approx$15000), yields a similar level of accuracy, with a typical increment of $\sigma_\rho$ of a mere factor 2-4.

To better appreciate the efficacy of the NN model, in Fig.~\ref{fig::size}(b) we superpose the $\rho(x)$ profiles predicted by each of the three NN models with the "true" one calculated by GA, for one of the Perlin-noise profiles in the test set. A substantial overlap is found for all three cases, with local differences that can noted only in the $\delta_\rho(x)$ plot as small uncorrelated fluctuations. Deviations decrease when taking larger datasets as also indicated by the $\sigma_\rho$ errors, still on the order of 10$^{-3}$.

There are however other cases in which the importance of building a large dataset becomes more evident, as for example the series of cusps reported in Fig.~\ref{fig::size}(c). The $N_\text{set}\approx$190000 NN model returns a one-to-one match of the GA $\rho(x)$ profile as indicated by the small $\delta_\rho(x)$ and $\sigma_\rho$. The $N_\text{set}\approx$70000 NN model still performs reasonably well but larger deviations are observed around the $\rho$ peaks and the $\sigma_\rho$ approaches the acceptance threshold. On the contrary, the NN model trained on the smallest $N_\text{set}\approx$15000 dataset does not match the actual peak values and returns visible oscillations in-between, ending in an above-threshold $\sigma_\rho$. It is worth noticing how, even in this worst case, the NN still provides a qualitative representation of the right features of the $\rho$ profile so that our strict 0.02 acceptance criterion looks appropriate to assess a quantitative match. For the following analysis, we will consider the most accurate $N_\text{set}\approx$190000 NN model in order to minimize any effect related to the dataset size.

\subsection{NN performance on sinusoidal and gaussian profiles}\label{sec::sin_gauss}
\begin{figure*}[ht!]
 \centering
   \includegraphics[width=\textwidth]{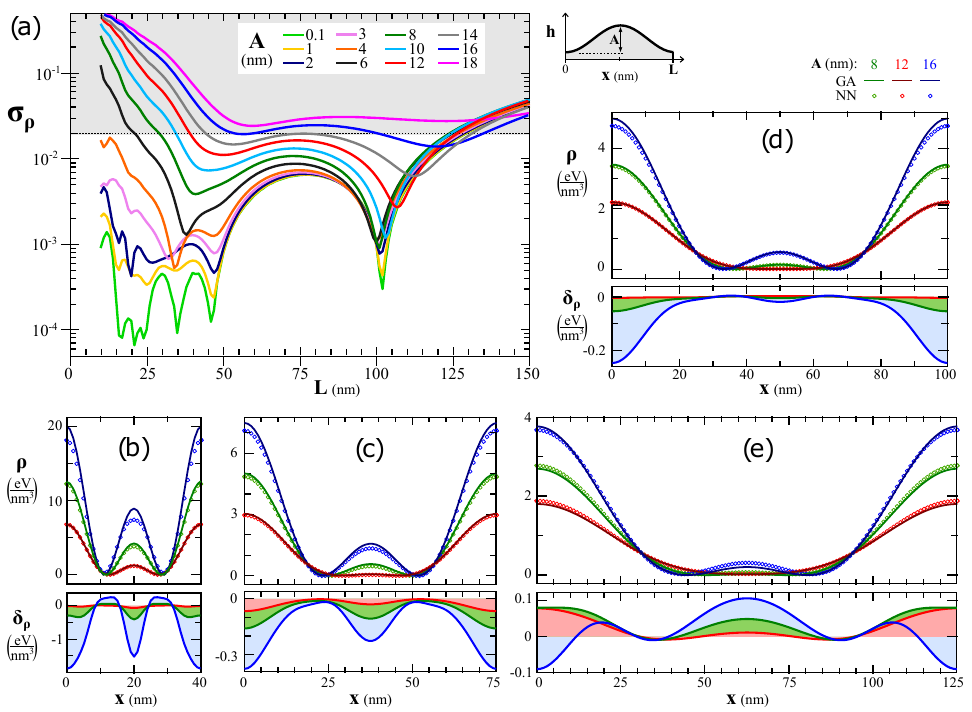}
    \caption{Analysis of the NN performances for single-period cosine profiles. (a) Plot of the $\sigma_\rho$ prediction error as a function of wavelength $\lambda\in[10,150]$~nm and height $A\in[0.1,18]$~nm. (b-e) Comparison between the NN prediction (circles) and the GA calculation (solid line) of $\rho$ for $\lambda=$40 (b), 75 (c), 100 (d) and 125~nm (e), and $A$=8, 12 and 16~nm. The local prediction error $\delta_\rho$ is reported for each case.}
    \label{fig::cosine}
\end{figure*}

To get a deeper insight into how the NN deals with arbitrary morphologies, we now inspect how it performs on cosine profiles of different wavelengths $\lambda$, i.e. on the single Fourier components that could form the spectrum of a generic function. For this analysis, we vary the domain size so to match a single cosine period, i.e. $L=\lambda$, while keeping the same space discretization step, $\delta x$, used in the training dataset. In Fig.~\ref{fig::cosine}(a) we report $\sigma_\rho$ as a function of the wavelength $\lambda$. $\lambda$ ranges from a minimum of 10~nm, imposed by the resolution limit of the $\delta x$=1~nm grid, to a maximum of 150~nm, exceeding the size of the training set (100~nm). This also tests the NN generalization capabilities to different domain sizes made possible by the Fully Convolutional architecture. Within this range, calculations are taken every 1~nm, allowing for the analysis of the interpolation behavior in between the few $\lambda$ components actually present in the fixed-length training domain, i.e. the integer sub-multiples of 100~nm. The error analysis is repeated on several cosine heights $A\in $[0.001, 20]~nm, which also extends beyond the maximum height (8~nm) of the profiles used in the training set. All cases used for this analysis are available from \cite{repository}.

As made evident in the plot, all interpolation cases for $\lambda\in$[10,100]~nm return errors well below the 0.02 acceptance threshold as long as the profile height $A$ remains lower than 4~nm. For taller profiles a loss in accuracy is observed, starting from smaller values of $\lambda$. At $A$=12~nm the NN still performs well down to $\lambda\sim$40~nm while for greater amplitudes the error increases and predictions remain reliable only in a narrow range of $\lambda$ around 100~nm, corresponding to the size of the domain used for training. This is made evident in the panels (b-d) of Fig.~\ref{fig::cosine} where the analytical and predicted $\rho$ profiles (top) and the corresponding local error $\delta_\rho$ (bottom) are superimposed for cosines of $\lambda$=40~nm (b), 75~nm (c) and 100~nm (d), and three different profile heights $A$=8, 12, and 16~nm. It is worth noticing that, even in the worst case of $\lambda$=40~nm and $A$=16~nm, corresponding to an above-threshold $\sigma_\rho\approx$0.06, the predicted $\rho$ is still qualitatively consistent with the analytical one despite the tendency to underestimate the actual variation range. A clear modulation of the $\sigma_\rho$ error can also be recognized in the plot of panel (a), indicating better predictivity for those $\lambda$s which are close to integer sub-multiples of 100~nm, i.e. those represented in the training dataset. The errors obtained for these cases are the same if repeating the cosines within a $L$=100~nm domain as used in training, given the inherent periodicity of our NN definition. For the intermediate $\lambda$s, $\sigma_\rho$ increases by about one order of magnitude which can be considered as the NN interpolation error. Finally, when considering cosines with $\lambda>$100~nm, i.e. exceeding the size of the training domain, a progressive growth of the $\sigma_\rho$ error is found, hitting the acceptance threshold at a maximum $\lambda\approx$125~nm for low-amplitude profiles. To better appreciate the quality of the generalization, the NN and GA profiles of $\rho$ for a $\lambda$=125~nm cosine are compared in Fig.~\ref{fig::cosine} (e) for the three different heights $A$=8, 12, and 16~nm. The agreement is overall satisfactory as all three cases correspond to near-threshold $\sigma_\rho$. As for the other cases in panels (b,d), major discrepancies are at the cosine top and bottom, as the NN tends to overestimate the $\rho$ values.

\begin{figure}[t!]
 \centering
   \includegraphics[width=\columnwidth]{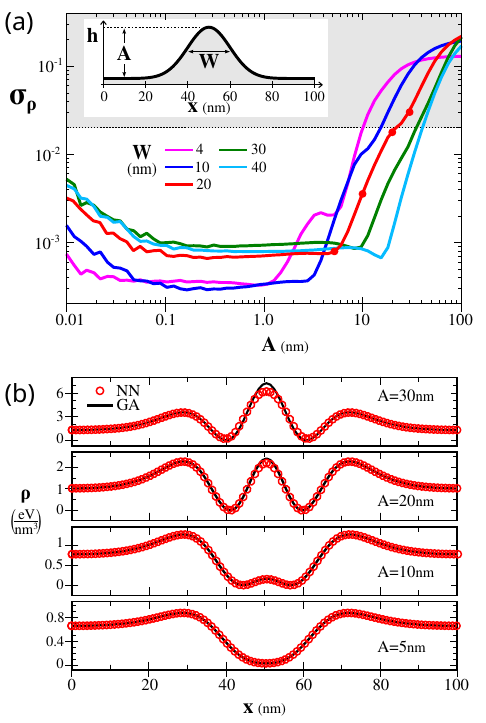}
    \caption{Analysis of the NN performances for gaussian profiles of different height $A$ and full-width at half-height $W$, centered in a $L$=100~nm cell. (a) Plot of the $\sigma_\rho$ prediction error as a function of the peak height $A$ from 0.01 to 100~nm for $W\in$ [4,40]~nm. (b-e) Superposition of $\rho$ profiles predicted by the NN and calculated by GA for a gaussian peak of $W$=20~nm and different heights $A$=5, 10, 20, and 30~nm, marked by dots in the plot of panel (a).}
    \label{fig::gaussian}
\end{figure}

The analysis of the NN performance on single Fourier modes provides some insights on its potential for predicting generic profiles but, since the NN is non-linear, it is not possible to assess its actual behavior on complex profiles by a mere Fourier synthesis approach. It is indeed expected that the NN performs better on the overall profile than on its single Fourier components, as the training did not include sinusoids. For this reason, we also inspected the NN predictivity on more localized peaks set by gaussians of width $W<$100~nm, whose spectrum would return important contributions from short wavelengths. Fig.~\ref{fig::gaussian}(a) reports the prediction error $\sigma_\rho$ for gaussian peaks of different widths $W$ in the range [4,40]~nm, at the center of a $L$=100~nm domain (see inset of Fig.~\ref{fig::gaussian}(a)), as a function of the peak height $A$. The NN matches the analytical solution, with $\sigma_\rho<$0.02, as far as $A$ remains below a critical height, proportional to the gaussian width $W$, i.e. as far as the profile aspect ratio remains on the order of the typical features included in the training set. The closer inspection of the case with $W$=20~nm in Fig. \ref{fig::gaussian}(b) reveals that for the tallest peaks the NN fails in reproducing the elastic response at the gaussian top, with the same systematic underestimation of $\rho^\text{GA}$ already observed for the taller cosine profiles (Fig.~\ref{fig::cosine}(b,c)).

\subsection{Large domain generalization}\label{sec::large_domain_statics}

\begin{figure*}[htb!]
 \centering
   \includegraphics[width=\textwidth]{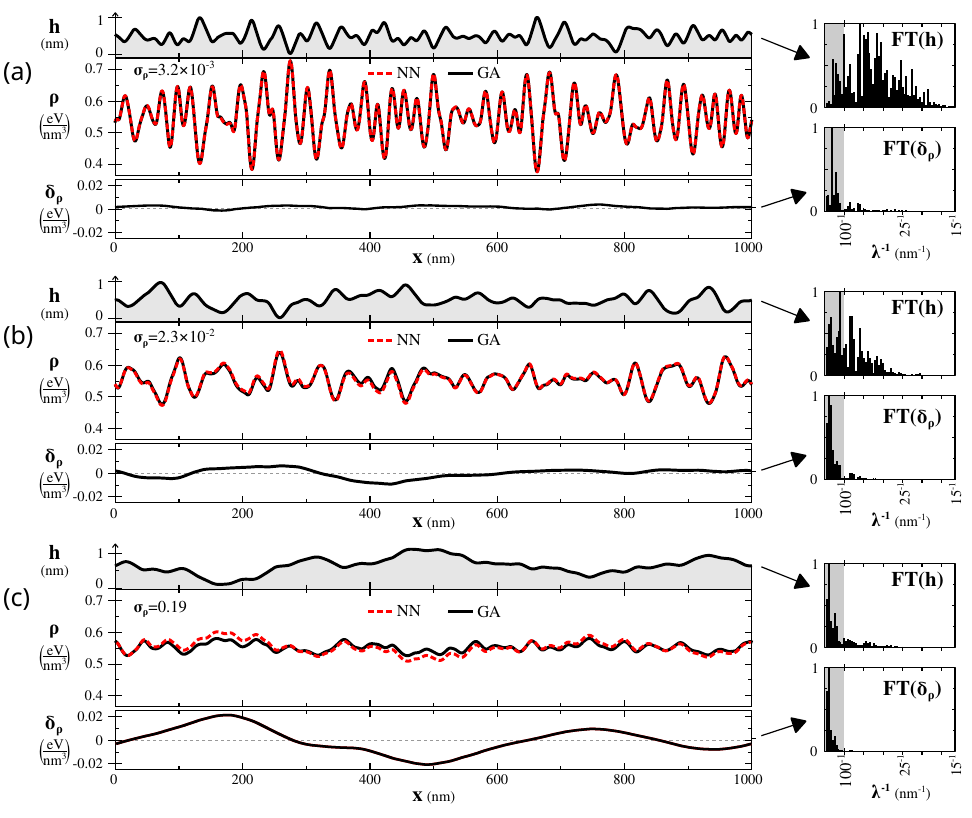}
    \caption{Analysis of the NN performance on a large ($L$=1000~nm) domain for three profiles $h(x)$ characterized by leading Fourier components (a) mostly within the $L$=100~nm training domain; (b) both shorter and longer than 100~nm; (c) longer than the 100~nm training size. For each profile, the NN predicted $\rho$ is plotted (red line) superimposed to the GA calculated one (black) line along with their difference $\delta_\rho=\rho^\text{NN}-\rho^\text{GA}$. The Fourier spectrum of each surface profile $\text{FT}(h)$ and that of the local prediction error $\text{FT}(\delta_\rho)$ are also reported for each case.}
    \label{fig::extended}
\end{figure*}

Thanks to the Fully Convolutional architecture of the NN, the trained model can be evaluated on any domain size irrespective of the one used for the training. Fig.~\ref{fig::cosine} already shows that NN results on single Fourier components are accurate up to an extension in domain size of $\sim$25\% beyond the one of the training set. Here, we want to test how the NN performs on generic profiles on a domain 10 times larger than the training one. This is illustrated in Fig.~\ref{fig::extended} where NN predictions are compared to GA calculations of $\rho$ for three surface profiles with different Fourier spectrum: case (a) is chosen so that it contains only Fourier components of $\lambda$s that mostly fall within the 100~nm length of the training set; case (b) combines both $\lambda$ lower and larger than 100~nm while in case (c) the leading $\lambda$s are all above the 100~nm training size. As expected, the NN provides a quantitative prediction of $\rho$ in case (a) with a prediction error $\sigma_{\rho}$ in the same order of $10^{-3}$ found for the testing cases. In case (b), the agreement is still overall satisfactory but some discrepancies appear at the maxima and minima of $\rho$, returning a $\sigma_\rho$ error around the 0.02 threshold. The prediction for case (c), instead, exhibits a $\sigma_\rho$ value above such threshold. This suggests that the prediction is only qualitative: the NN captures the position of $\rho$ peaks but cannot match their absolute values. As we already pointed out, the NN architecture is non-linear so the failure in the prediction of long-wavelength is not a straightforward consequence of a Fourier decomposition. A more in-depth analysis, however, reveals that NN inaccuracies are mainly related to long-range effects. This is actually expected since the elastic interaction is long-range by definition while the NN training on a finite domain size is equivalent to introducing a cutoff distance. Fourier analysis of the local prediction error $\delta_\rho$ clearly reveals that the discrepancies are modulated by long-$\lambda$ modes only, where the NN was not trained. This suggests that the NN can be proficiently applied to larger domain sizes whenever the physical scale of the features appearing on the surface is comparable to the training set one. If the profile features contain longer wavelengths, on the other hand, the NN predictions will match the real ones only up to long-range contributions. While the overall prediction might not be quantitative, this means that the relative values of elastic energy density in neighboring positions are still trustworthy.

\subsection{Surface dynamics using the deep-learned $\rho$ as driving force}\label{sec::dynamics}
In the previous sections, we showed how the NN approach can be exploited to extract the map of elastic energy density $\rho(x)$ for a generic profile $h(x)$, matching the same level of accuracy of the analytical calculation in the appropriate parameter range. In this section, we further test the robustness of such predictions when used for the numerical integration of time-dependent partial differential equations \cite{Mullins1957JAP,FreundBOOK2003,fried2004unified} describing the evolution of the surface morphology $h(x,t)$ of the strained film.

\begin{figure*}[htb!]
 \centering
   \includegraphics[width=\textwidth]{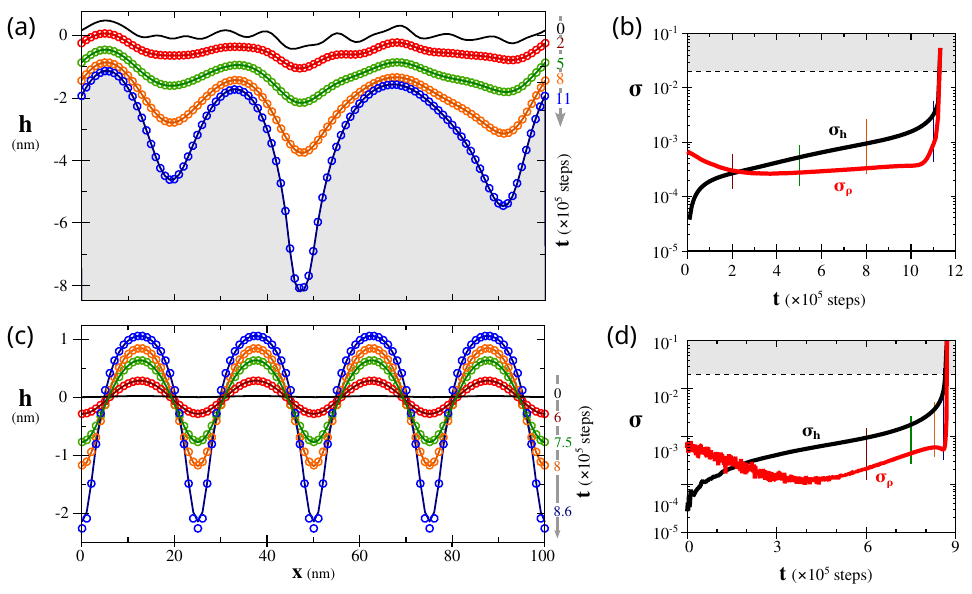}
    \caption{Comparison between evolution profiles obtained by time-integration using both the GA and the NN evaluations of the elastic contribution $\rho$ to the surface chemical potential for: (a) evaporation dynamics with $k$=1 ($k\delta t$=10$^{-5}$~a.u.) and $\mu_0$=0.7 eV/nm$^3$ for an initial Perlin noise profile; (c) ATG-like dynamics starting from a $\lambda$=25~nm cosine of amplitude 0.01~nm ($D\delta t$=5$\times$10$^{-4}$ a.u.). The respective variations of the prediction error $\sigma$ for $h(x)$ and $\rho$ as a function of time $t$, are shown in panels (b) and (d).}
    \label{fig::dynamics}
\end{figure*}

In particular, we consider two different regimes: 
\begin{enumerate}[(a)]
   \item the evaporation/condensation dynamics, where the profile directly moves according to the difference of the local chemical potential $\mu$ with respect to a reference value $\mu_0$, e.g. the one of the gas phase:
   \begin{equation}\label{eq::evap}
     \frac{dh}{dt} = -k\sqrt{1+h'^2} (\mu-\mu_0)\,,
   \end{equation}
   with $k$ a kinetic coefficient; 

   \item the surface diffusion dynamics where material flows along the surface are set by the gradients of $\mu$:  
   \begin{equation}\label{eq::diff}
     \frac{dh}{dt} = D\sqrt{1+h'^2} \frac{\partial^2\mu}{\partial s^2}\,,
   \end{equation}
   with $D$ the diffusion coefficient and $s$ the surface coordinate. 
\end{enumerate}
In its simplest formulation, $\mu$ comprises a (isotropic) surface energy contribution, proportional to the local curvature $\kappa=-h''(1+h'^2)^{-3/2}$, and an elastic energy contribution which is directly given by the elastic energy density $\rho$:
\begin{equation}
    \mu=\kappa\gamma+\rho\,,
\end{equation}
where $\gamma$ is the surface energy density, here set equal to 60~meV/\AA{}$^2$ to mimic Ge. The time-integration of these evolution equations is numerically performed by a simple Euler explicit scheme with a fixed timestep $\delta t$. For convenience, the simulation time $t$ is counted as the number of timesteps, scalable to the physical values by the choice of $k\delta t$ or $D\delta t$.

The NN performance is then evaluated by comparing the profile evolution obtained by using the predicted $\rho$ in the time integration with the ground-truth one obtained by using its explicit GA calculation. Quantitatively, we monitor both the error $\sigma_\rho$ (eq.~\eqref{eq::nrmseeq}) on the prediction of the $\rho$ profile at each time step and the error $\sigma_h$ accounting for the difference between the $h^\text{NN}$ profile obtained at a certain time $t$ when using the NN predictions of $\rho$ instead of the true one $h^\text{GA}$ relying on the GA calculation of $\rho$ at each timestep:
\begin{equation}
    \sigma_h(t) = \frac{\sqrt{\dfrac{1}{N_x}\sum_i^{N_x} \left[h^\text{NN}(x_i,t) - h^\text{GA}(x_i,t)\right]^2}}{\text{max}(h^\text{GA})-\text{min}(h^\text{GA})} \,.
\end{equation}

In Fig.~\ref{fig::dynamics} we show two representative examples of surface evolution simulations, demonstrating the efficacy of the NN approach. Several additional simulation cases are available in the repository \cite{repository}.

The first case in panel (a) shows a simulation of surface evaporation according to eq.~\eqref{eq::evap}. As the evaporation rate is not uniform along the profile, a general coarsening of the Perlin-noise features occurs while deep trenches develop due to the local strain concentration. Despite such divergent dynamics, no distinguishable difference can be observed between the NN-based profiles and the GA ones over the whole time integration range. A more quantitative analysis of the NN accuracy is provided in panel (b) where both $\sigma_\rho$ and $\sigma_h$ prediction errors are traced over time. As expected from Sect.~\ref{sec::results_trainvalidtest}, the error $\sigma_\rho$ on the NN estimation of $\rho$ remains well below the acceptance threshold during the whole evolution but for the last divergent stages. The slow increment of $\sigma_h$ shows that the error accumulation due to the use of $\rho^\text{NN}$ during the time integration is not critical to follow the ``true" GA dynamics. 

In Fig.~\ref{fig::dynamics}(c) we instead consider a simulation by surface diffusion dynamics (eq.~\eqref{eq::diff}) of ATG-like film instability, initiated from a cosine perturbation of unstable wavelength $\lambda$=25~nm and height $A_0$=0.02~nm. A substantial match is recognizable in all reported stages, covering both the early linear regime (up to $t\approx$5$\times$10$^5$ steps) and the subsequent non-linear evolution toward cusp shapes, eventually ending in a numerical divergence as the cusp points sharpen (at $t\approx$8.7$\times$10$^5$ steps). Small discrepancies are just observed at the latest evolution stages, in correspondence of such critical points. The analysis of the prediction errors $\sigma_\rho$ and $\sigma_h$ reported in panel (d), confirms the one-to-one match, with both indicators remaining below the 0.02 threshold at any time, with limited accumulation effect in $\sigma_h$. 

The one-to-one match found for both (a) and (c) cases of Fig.~\ref{fig::dynamics}, despite their exponential behavior spanning a variation of 3 orders of magnitude in profile height, is a strong indication of reliability.

\begin{figure*}[t!]
 \centering
   \includegraphics[width=\textwidth]{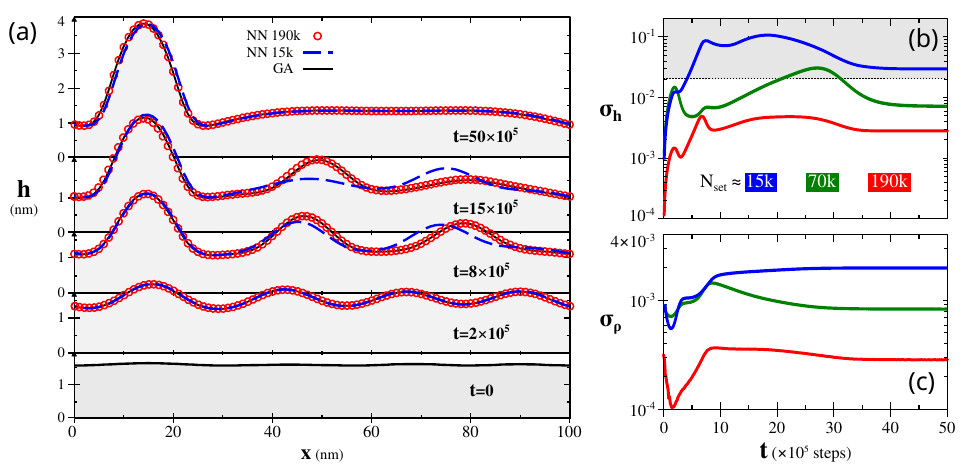}
    \caption{Simulation of island formation and coarsening for a Ge film on a Si substrate (at $x<0$) by surface diffusion dynamics ($D\delta t$=5$\times$10$^{-3}$ a.u.), including a wetting energy contribution to $\mu$. The initial profile is modulated by four cosine peaks of different $\lambda$s and amplitude $A\in[0.02,0.065]$~nm and has an unstable thickness $\left\langle h \right\rangle\approx$1.5~nm.
    (a) Evolution profiles obtained by using both the GA and NN evaluations of $\rho$ in the time-integration. The predictions from the two NN models for $N_\text{set}\approx$15000 and 190000 are compared. The corresponding variation of the prediction error $\sigma$ for $h(x)$ and $\rho$ as a function of time $t$, are shown in panels (b) and (c). The same curves for the $N_\text{set}\approx$70000 NN model are also reported.}
    \label{fig::islandsize}
\end{figure*}

Since the NN only copes with the elastic energy contribution to $\mu$, other terms, not related to strain relaxation, can be easily plugged in without the need to re-train a dedicated model. We then consider a more realistic description of a thin Ge film on a Si substrate by including a wetting energy contribution to the chemical potential \cite{BergamaschiniAPX2016}, accounting for the exponential decay of the film surface energy density $\gamma$ as a function of its thickness \cite{LuPRL2005}. Following Ref.~\cite{PhysRevB.76.165319}, we define $\gamma(h)=\gamma_\text{Ge}+(\gamma_\text{Si}-\gamma_\text{Ge})\exp(-h/d)$, with $\gamma_{Si}$=8.7~eV/nm$^2$ and $d$=0.27~nm. The result of such a contribution is the stabilization of the flat profile for a thickness below a critical one. In Fig.~\ref{fig::islandsize}(a) we show the morphological evolution of a slightly-perturbed, planar film of thickness right above the critical value, as obtained by using $\rho$ from the $N_\text{set}\approx$190000 and 15000 NN models with respect to using the true one by GA in the time integration of the surface diffusion equation. The early stages of the evolution follow ATG dynamics, with the perturbation amplitude growing almost exponentially in time ($t<$2$\times$ 10$^5$ steps). When large enough trenches start to be dug, the wetting term becomes relevant and locally quenches the dynamics, so that the profile consists of separate "islands" sitting on top of a wetting layer, as for Stranski-Krastanov conditions \cite{FreundBOOK2003}. Then, slow coarsening occurs with lower peaks being consumed by the largest, more stable, ones ending in a stationary state ($t\approx$50$\times$10$^5$ steps) made of a single island on top of a flat wetting layer. While this general behavior is obtained for both NN models, only the one trained on the full dataset is capable of following one-to-one the GA evolution at all evolution stages. In contrast, the $N_\text{set}\approx$15000 NN model fails in reproducing the actual geometry and coarsening of peaks, resulting in noteworthy discrepancies that are partly healed in the late stages as approaching a stationary state consistent with the GA one. For a more quantitative analysis, in panels (b) and (c) we monitor the variation of both $\sigma_h$ and $\sigma_\rho$ errors over time, for both the $N_\text{set}\approx$190000 and 15000 NN models considered in panel (a) and for the mid-sized model for $N_\text{set}\approx$70000. All models are found to yield appropriate predictions of $\rho$ at all evolution stages as the $\sigma_\rho$ remains at least one order of magnitude lower than the 0.02 acceptance threshold. Given the stabilizing nature of the wetting contribution introduced in the dynamics, the $\sigma_h$ error does not grow continuously as in the cases of Fig.~\ref{fig::dynamics} but shows maxima at the onset of each island coarsening and then tends to stabilize in the subsequent evolution steps. In the case of the largest dataset the small $\sigma_\rho$ error, always of the order of 10$^{-3}$ confirms the quantitative reproduction of the GA evolution at any step. A satisfactory comparison is also obtained for the intermediate size, with a limited time interval where the error grows above the threshold. For the smallest dataset, we instead see that $\sigma_h$ quickly grows above the threshold and then the predicted evolution cannot be taken as a perfect replacement of the true GA one. By this analysis, we can then conclude that the capability of providing a good, below-threshold prediction of the static $\rho$ profile is not sufficient to guarantee a stable and consistent description of the system dynamics due to such a term. However, the larger is the dataset used for the NN training, the lower will be the error propagation across time iterations, thus returning more reliable predictions even after millions of integration steps.

\begin{figure*}[t!]
 \centering
   \includegraphics[width=\textwidth]{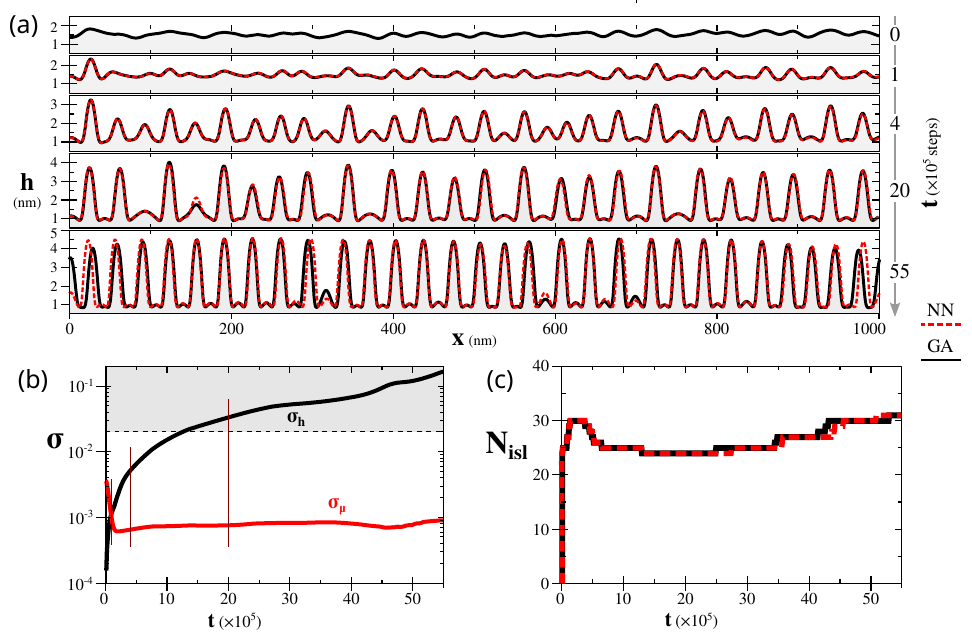}
    \caption{Analysis of the NN performance in large-scale ($L=1000$~nm) and long-time simulations. (a) Superposition of GA and NN evolution profiles for a simulation of island growth due to surface diffusion ($D\delta t$=5$\times$10$^{-3}$) and deposition by a constant vertical flux ($D/f$=5$\times$10$^4$). (b) Evolution of the $\sigma_\rho$ and $\sigma_h$ prediction errors. (c) Variation of the number of islands $N_\text{isl}$ for both GA and NN evolutions.}
    \label{fig::dynamicsLarge}
\end{figure*}

Given the extrapolation capabilities of the NN demonstrated in Figure~\ref{fig::extended} it is worth investigating if they hold true also for the dynamics on large domains and long simulation times. In Fig.~\ref{fig::dynamicsLarge} we analyze the performance of the NN approach when applied to the simulation of large-scale and long-time evolutions. In particular, we consider a $L$=1000~nm domain, discretized with the same resolution $\delta x$=1~nm considered so far, and initialize a surface profile by random Fourier components with $\lambda \le$100~nm for which the NN was demonstrated (Fig.~\ref{fig::extended}(a)) to return good prediction of $\rho$. The same islanding dynamics of Figure~\ref{fig::islandsize}(a) is considered but for the addition of a deposition flux, incrementing the film thickness by a constant rate $f$. The overall behavior is analogous to the description of Fig.~\ref{fig::islandsize}(a) but for the prolonged stability of pristine islands due to the continuous material supply by the deposition. Indeed, all peaks with $\lambda\approx\lambda_\text{max}$ grow at larger aspect ratios and start coarsening only in the latest evolution stages until divergent cusp points appear. The NN is found to match the GA dynamics during the earlier growth stages where the initial perturbation grows in amplitude. Instead, local discrepancies emerge in the subsequent stages whenever a peak is extinguished or a new one is formed. These faults should not be ascribed to a loss in the reliability of the NN prediction of $\rho$, which indeed is characterized by $\sigma_\rho$ well below the acceptance threshold at all simulation times (see Fig.~\ref{fig::dynamicsLarge}(b)), but to the inherent instability of the coarsening dynamics, sensitive to whatever small differences between the peaks. Consequently, $\sigma_h$ is found to monotonously increase over time up to about 0.15 at the latest evolution stages where a few peaks are missing (see e.g. the peaks at $x\approx$0 or 580~nm for the profile at $t$=55$\times$10$^5$ steps in panel (a)). At this point, the NN does not provide a one-to-one replacement of the analytic GA calculation. Nonetheless, on a qualitative ground, the NN evolution still looks consistent with the GA one and hence it can still be exploited for physically meaningful investigations on the overall trends and properties. As an example, in Fig.~\ref{fig::dynamicsLarge}(c) we monitor the variation in time of the number of islands $N_\text{isl}$, i.e. the number of peaks above a threshold quota ($h_c$=1.5~nm). A substantial agreement is found between the curve obtained from the NN-based simulation and the one based on GA with limited time shifts corresponding to a lack of synchronization in the coarsening of single peaks. Such deviations are expected to have negligible effects when investigating the average system behavior based on properly averaging over multiple cases. 

\section{Conclusions}
In this work, we showed how a specialized Machine Learning approach can be leveraged to quantitatively predict the elastic energy density, $\rho$, at the surface of a Ge/Si(001) strained film, surrogating its explicit calculation here based on a semi-analytical Green's function approximation. 

A large dataset of about 190000 examples has been generated and three NN models have been trained, one using the entire set and the others taking a fraction of $\approx$37\% and $\approx$8\% of it, so to inspect the effects and trade-offs on the resulting NN predictions. All models converged during training and demonstrated quantitative predictions of $\rho$ on the validation set and on additional cases of new classes of profiles, with better accuracy and generalization performances for the larger datasets. Still, results suggest that even a moderate number of examples ($\approx$~15000) can be used to train NN models with usable outputs, especially for qualitative assessment. Accuracy becomes instead critical if NN predictions are to be used in multiple iterations, as for the time integration of surface evolution equations, because of the error accumulation so that only the models trained on a sufficiently large dataset can be trustable.

As the ML architecture is based on a Fully Convolutional NN, it can be used on computational cells larger than those used in training. Limitations of the generalization capabilities have also been quantitatively investigated in this case. The NN model exhibits quantitative agreement as long as the typical wavelength of profile features is comparable to the size of the training set. If, instead, long-range contributions are critical, a loss in prediction accuracy is observed. This behavior is expected due to the inherent non-locality of the elastic equilibrium solution. Future development of ML architectures for this class of applications should therefore take into account this problem using global free surface descriptors or alternative NN modules (e.g. attention mechanism, Fourier neural operators~\cite{vaswani2017attention, li2020fourier}).

While the present work takes profit from the simplicity and low-computational cost of Green's approximation method to construct arbitrary large datasets and have available the ground-truth $\rho$ profiles for any analysis, its limitation to small-slope profiles hinders the application to more realistic cases. Moreover, no practical advantage comes out of replacing GA with NN as the actual speed-up is negligible. The main achievement of this study is then to provide a comprehensive analysis of the NN approach to be transferred to other, more advanced calculation techniques such as Finite Element Method for which a NN surrogate represents a significant cut in computational cost, as will be shown in an upcoming work.

\section*{Acknoledgements}
R.B. and F.M. acknowledge financial support from ICSC – Centro Nazionale di Ricerca in High Performance Computing, Big Data and Quantum Computing, funded by European Union – NextGenerationEU”. L.M. acknowledges the support offered by MCIN/AEI/ 10.13039/501100011033 under Project No. PID2020-115118GB-I00.

\bibliography{references.bib}

\begin{thebibliography}{10}

\bibitem{goodfellowdeep2016}
I.~Goodfellow, Y.~Bengio, and A.~Courville, {\em Deep Learning}.
\newblock MIT Press, 2016.
\newblock \url{http://www.deeplearningbook.org}.

\bibitem{Mehta2019PhysRep}
P.~Mehta, M.~Bukov, C.~H. Wang, A.~G. Day, C.~Richardson, C.~K. Fisher, and
  D.~J. Schwab, ``{A high-bias, low-variance introduction to Machine Learning
  for physicists},'' {\em Physics Reports}, vol.~810, pp.~1--124, 2019.

\bibitem{Behler2007PRL}
J.~Behler and M.~Parrinello, ``{Generalized neural-network representation of
  high-dimensional potential-energy surfaces},'' {\em Physical Review Letters},
  vol.~98, pp.~1--4, 2007.

\bibitem{Bartok2010PRL}
A.~P. Bart{\'{o}}k, M.~P. Payne, R.~Kondor, and G.~Cs{\'{a}}nyi, ``{Gaussian
  Approximation Potentials: The Accuracy of Quantum Mechanics, without the
  Electrons},'' {\em Physical Review Letters}, vol.~104, p.~136403, 2010.

\bibitem{venderley2018machine}
J.~Venderley, V.~Khemani, and E.-A. Kim, ``Machine learning out-of-equilibrium
  phases of matter,'' {\em Physical review letters}, vol.~120, no.~25,
  p.~257204, 2018.

\bibitem{salmenjoki_machine_2018}
H.~Salmenjoki, M.~J. Alava, and L.~Laurson, ``Machine learning plastic
  deformation of crystals,'' {\em Nature Communications}, vol.~9, p.~5307,
  2018.

\bibitem{Jha2018grain_orientation}
D.~Jha, S.~Singh, R.~Al-Bahrani, W.-k. Liao, A.~Choudhary, M.~De~Graef, and
  A.~Agrawal, ``{Extracting Grain Orientations from EBSD Patterns of
  Polycrystalline Materials Using Convolutional Neural Networks},'' {\em
  Microscopy and Microanalysis}, vol.~24, pp.~497--502, 10 2018.

\bibitem{Yang2021dopants}
S.-H. Yang, W.~Choi, B.~W. Cho, F.~O.-T. Agyapong-Fordjour, S.~Park, S.~J. Yun,
  H.-J. Kim, Y.-K. Han, Y.~H. Lee, K.~K. Kim, and Y.-M. Kim, ``Deep
  learning-assisted quantification of atomic dopants and defects in 2d
  materials,'' {\em Advanced Science}, vol.~8, no.~16, p.~2101099, 2021.

\bibitem{ge2024silicon}
G.~Ge, F.~Rovaris, D.~Lanzoni, L.~Barbisan, X.~Tang, L.~Miglio, A.~Marzegalli,
  E.~Scalise, and F.~Montalenti, ``Silicon phase transitions in
  nanoindentation: Advanced molecular dynamics simulations with machine
  learning phase recognition,'' {\em Acta Materialia}, vol.~263, p.~119465,
  2024.

\bibitem{cybenko1989approximation}
G.~Cybenko, ``Approximation by superpositions of a sigmoidal function,'' {\em
  Mathematics of control, signals and systems}, vol.~2, no.~4, pp.~303--314,
  1989.

\bibitem{raissi2019physics}
M.~Raissi, P.~Perdikaris, and G.~E. Karniadakis, ``Physics-informed neural
  networks: A deep learning framework for solving forward and inverse problems
  involving nonlinear partial differential equations,'' {\em Journal of
  Computational physics}, vol.~378, pp.~686--707, 2019.

\bibitem{Fulton2019CGF}
L.~Fulton, V.~Modi, D.~Duvenaud, D.~I. Levin, and A.~Jacobson, ``{Latent-space
  Dynamics for Reduced Deformable Simulation},'' {\em Computer Graphics Forum},
  vol.~38, pp.~379--391, 2019.

\bibitem{MontesdeOcaZapiain2021npj}
D.~{Montes de Oca Zapiain}, J.~A. Stewart, and R.~Dingreville, ``{Accelerating
  phase-field-based microstructure evolution predictions via surrogate models
  trained by machine learning methods},'' {\em npj Computational Materials},
  vol.~7, pp.~1--11, 2021.

\bibitem{Yang2021Patterns}
K.~Yang, Y.~Cao, Y.~Zhang, S.~Fan, M.~Tang, D.~Aberg, B.~Sadigh, and F.~Zhou,
  ``{Self-supervised learning and prediction of microstructure evolution with
  convolutional recurrent neural networks},'' {\em Patterns}, vol.~2,
  p.~100243, 2021.

\bibitem{Lanzoni2022PRM}
D.~Lanzoni, M.~Albani, R.~Bergamaschini, and F.~Montalenti, ``Morphological
  evolution via surface diffusion learned by convolutional, recurrent neural
  networks: Extrapolation and prediction uncertainty,'' {\em Physical Review
  Materials}, vol.~6, p.~103801, 2022.

\bibitem{Zeqing2023APLML}
Z.~Jin, B.~Zheng, C.~Kim, and G.~X. Gu, ``{Leveraging graph neural networks and
  neural operator techniques for high-fidelity mesh-based physics
  simulations},'' {\em APL Machine Learning}, vol.~1, p.~046109, 11 2023.

\bibitem{landau}
L.~D. Landau and E.~M. Lifshitz, {\em Theory of Elasticity}.
\newblock Butterworth-Heinemann, Elsevier, 1986.

\bibitem{tersoff1994}
J.~Tersoff and F.~K. LeGoues, ``Competing relaxation mechanisms in strained
  layers,'' {\em Physical Review Letters}, vol.~72, pp.~3570--3573, 1994.

\bibitem{Zinovyev}
V.~Zinovyev, G.~Vastola, F.~Montalenti, and L.~Miglio, ``Accurate and
  analytical strain mapping at the surface of ge/si(001) islands by an improved
  flat-island approximation,'' {\em Surface Science}, vol.~600, p.~4777–4784,
  2006.

\bibitem{long2015fully}
J.~Long, E.~Shelhamer, and T.~Darrell, ``Fully convolutional networks for
  semantic segmentation,'' in {\em Proceedings of the IEEE conference on
  computer vision and pattern recognition}, pp.~3431--3440, 2015.

\bibitem{FreundBOOK2003}
L.~B. Freund and S.~Suresh, {\em Thin Film Materials: Stress, Defect Formation
  and Surface Evolution}.
\newblock Cambridge University Press, 2004.

\bibitem{Mullins1957JAP}
W.~W. Mullins, ``{Theory of Thermal Grooving},'' {\em Journal of Applied
  Physics}, vol.~28, p.~333, 1957.

\bibitem{BergamaschiniAPX2016}
R.~Bergamaschini, M.~Salvalaglio, R.~Backofen, A.~Voigt, and F.~Montalenti,
  ``Continuum modelling of semiconductor heteroepitaxy: an applied
  perspective,'' {\em Advances in Physics: X}, vol.~1, pp.~331--367, 2016.

\bibitem{eshelby1957determination}
J.~D. Eshelby, ``The determination of the elastic field of an ellipsoidal
  inclusion, and related problems,'' {\em Proceedings of the Royal Society of
  London. A}, vol.~241, pp.~376--396, 1957.

\bibitem{mura2013micromechanics}
T.~Mura, {\em Micromechanics of defects in solids}.
\newblock Springer Dordrecht, 2012.

\bibitem{paszkepytorch2019}
A.~Paszke, S.~Gross, F.~Massa, A.~Lerer, J.~Bradbury, G.~Chanan, T.~Killeen,
  Z.~Lin, N.~Gimelshein, L.~Antiga, A.~Desmaison, A.~Köpf, E.~Yang, Z.~DeVito,
  M.~Raison, A.~Tejani, S.~Chilamkurthy, B.~Steiner, L.~Fang, J.~Bai, and
  S.~Chintala, ``Pytorch: An imperative style, high-performance deep learning
  library,'' 2019.

\bibitem{choy2019fully}
C.~Choy, J.~Park, and V.~Koltun, ``Fully convolutional geometric features,'' in
  {\em Proceedings of the IEEE/CVF international conference on computer
  vision}, pp.~8958--8966, 2019.

\bibitem{lata2019image}
K.~Lata, M.~Dave, and K.~Nishanth, ``Image-to-image translation using
  generative adversarial network,'' in {\em 2019 3rd international conference
  on electronics, communication and aerospace technology (ICECA)},
  pp.~186--189, IEEE, 2019.

\bibitem{Schubert2019IEEE}
S.~Schubert, P.~Neubert, J.~Pöschmann, and P.~Protzel, ``Circular
  convolutional neural networks for panoramic images and laser data,'' in {\em
  2019 IEEE Intelligent Vehicles Symposium (IV)}, pp.~653--660, 2019.

\bibitem{Kayhan2020CVF}
O.~Semih~Kayhan and J.~C. van Gemert, ``On translation invariance in cnns:
  Convolutional layers can exploit absolute spatial location,'' in {\em 2020
  IEEE/CVF Conference on Computer Vision and Pattern Recognition (CVPR)},
  pp.~14262--14273, 2020.

\bibitem{Perlin1985287}
K.~Perlin, ``An image synthesizer,'' {\em SIGGRAPH Computer Graphics}, vol.~19,
  p.~287 – 296, 1985.

\bibitem{PythonPerlinNoise}
``Python implementation for perlin noise.''
  \url{https://pypi.org/project/perlin-noise/ }, 2023.

\bibitem{Asaro1972MetalTrans}
R.~J. Asaro and W.~A. Tiller, ``Interface morphology development during stress
  corrosion cracking: Part i. via surface diffusion,'' {\em Metallurgical
  Transactions}, vol.~3, pp.~1789--1796, 1972.

\bibitem{Grinfel1986SPD}
M.~Grinfeld, ``Instability of the separation boundary between a
  nonhydrostatically stressed elastic body and a melt,'' in {\em Soviet Physics
  Doklady}, vol.~31, p.~831, 1986.

\bibitem{Grinfeld1993JNS}
M.~Grinfeld, ``The stress driven instability in elastic crystals: Mathematical
  models and physical manifestations,'' {\em Journal of Nonlinear Science},
  vol.~3, pp.~35--83, 1993.

\bibitem{SrolovitzAM1989}
D.~Srolovitz, ``On the stability of surfaces of stressed solids,'' {\em Acta
  Metallurgica}, vol.~37, pp.~621--625, 1989.

\bibitem{repository}
L.~Martín-Encinar, D.~Lanzoni, A.~Fantasia, F.~Rovaris, R.~Bergamaschini, and
  F.~Montalenti, ``Deep learning of surface elastic chemical potential in
  strained films: from statics to dynamics,'' 2024.
\newblock {Materials Cloud Archive 2024.X}.

\bibitem{kingmaadam2014}
D.~P. Kingma and J.~Ba, ``Adam: A method for stochastic optimization,'' 2017.

\bibitem{bishop2006pattern}
C.~M. Bishop, {\em Pattern recognition and machine learning}.
\newblock Springer New York, NY, 2006.

\bibitem{fried2004unified}
E.~Fried and M.~E. Gurtin, ``A unified treatment of evolving interfaces
  accounting for small deformations and atomic transport with emphasis on
  grain-boundaries and epitaxy,'' {\em Advances in applied mechanics}, vol.~40,
  pp.~1--177, 2004.

\bibitem{LuPRL2005}
G.-H. Lu and F.~Liu, ``Towards quantitative understanding of formation and
  stability of ge hut islands on si(001),'' {\em Physical Review Letters},
  vol.~94, p.~176103, 2005.

\bibitem{PhysRevB.76.165319}
J.-N. Aqua, T.~Frisch, and A.~Verga, ``Nonlinear evolution of a morphological
  instability in a strained epitaxial film,'' {\em Physical Review B}, vol.~76,
  p.~165319, 2007.

\bibitem{vaswani2017attention}
A.~Vaswani, N.~Shazeer, N.~Parmar, J.~Uszkoreit, L.~Jones, A.~N. Gomez,
  {\L}.~Kaiser, and I.~Polosukhin, ``Attention is all you need,'' {\em Advances
  in neural information processing systems}, vol.~30, 2017.

\bibitem{li2020fourier}
Z.~Li, N.~Kovachki, K.~Azizzadenesheli, B.~Liu, K.~Bhattacharya, A.~Stuart, and
  A.~Anandkumar, ``Fourier neural operator for parametric partial differential
  equations,'' {\em arXiv preprint arXiv:2010.08895}, 2020.

\end{thebibliography}

\end{document}